\documentstyle[12pt,epsf,psfig]{article}
\begin{document}

\begin{center}
{\Large\bf 
Interplay between proton ordering and
ferroelectric polarization in H-bonded
KDP-type crystals} \\

\vspace{0.5cm}
{S. Koval,$^{\rm (1,2)}$ J. Kohanoff,$^{\rm (3,1)}$
 R.L. Migoni,$^{\rm (2,1)}$ and A. Bussmann-Holder $^{\rm (4)}$} \\

\vspace{0.6cm}
{\it $^{\rm (1)}$ International Centre for Theoretical Physics, \\ 
                  Strada Costiera 11, I-34014 Trieste, Italy }\\
{\it $^{\rm (2)}$ Instituto de F\'{\i}sica Rosario, Universidad
                  Nacional de Rosario, \\
                  27 de Febrero 210 Bis, 2000 Rosario, Argentina} \\
{\it $^{\rm (3)}$ Atomistic Simulation Group, Queen's University Belfast, \\
                  Belfast BT7 1NN, Northern Ireland}  \\
{\it $^{\rm (4)}$ Max-Planck-Institut f\"ur Festk\"orperforschung,
                  D-70506 Stuttgart, Germany} \\
\end{center}

\begin{abstract}
The origin of ferroelectricity in KH$_2$PO$_4$ (KDP) is studied by
first-principles electronic structure calculations. In the low-temperature
phase, the collective off-center ordering of the protons is accompanied by 
an electronic charge delocalization from the {\sl near} and localization at
the {\sl far} oxygen within the O-H$\cdots$O bonds. Electrostatic forces,
then, push the K$^+$ ions towards off-center positions, and induce a
macroscopic polarization. The analysis of the correlation between different
geometrical and electronic quantities, in connection with experimental data,
supports the idea that the role of tunnelling in isotopic effects is
irrelevant. Instead, geometrical quantum effects appear to play a central 
role.
\end{abstract}

\vspace{0.2cm}

\section{Introduction}

Among the hydrogen-bonded ferroelectrics, the compound
KH$_2$PO$_4$ (KDP) has been the most intensively
studied over the second half of the last century.\cite{blizec,lingla}
However, a detailed knowledge about
its ferroelectric phase transition and some related phenomena is
still lacking. In particular, its transition temperature (T$_c$ $\approx$
122K), shows an increase of about 107K upon deuteration. The origin of this
huge isotopic effect is still controversial. In the sixties, Blinc and Zecks
proposed a quantum tunnelling model to explain it, which focuses purely on
mass-dependent effects.\cite{Bli60,Bli72} Improvements of this model
include coupling to vibrations of the K-PO$_4$ complex.\cite{lingla}
Experimental studies over the past fifteen years, especially high-resolution 
neutron difraction measurements,\cite{Nel87,Ich87,Tun88,Nel88,McM90} 
provided increasing evidence that the geometrical modification of the 
hydrogen bonds upon deuteration, the well-known Ubbelohde effect,\cite{Ubb39} 
is intimately connected with the mechanism of the phase transition. 
McMahon {\it et al.} have shown that the distance $\delta$ between
the two collective equilibrium positions of the protons is remarkably 
correlated with T$_c$.\cite{McM90} They proposed that $\delta$ might be 
the only relevant parameter for T$_c$ and, in addition, no concluding 
evidence of tunnelling was established. It was then suggested 
the necessity of a revised theory for this class of transitions. Recently, 
a phenomenological model was proposed, which couples, via geometrical 
details of the H-bond, the proton coordinate with a nonlinear dynamics 
of the K-PO$_4$ system, where the spontaneous polarization actually 
develops.\cite{Bus98} This model, which includes both tunnelling and 
geometrical isotope effects, also allows to explain the T$_c$ behaviour 
upon deuteration.

Summarizing, it appears well-established that, in this class of systems, 
proton ordering and geometrical and electronic properties of the host 
cage are connected in a non trivial way.\cite{Kru90} So far, to the best
of our knowledge, no {\it ab initio} calculations have been reported for 
KDP. This type of approach has the advantage of allowing a confident
parameter-free analysis of the microscopic electronic and structural changes
upon proton ordering that drive the ferroelectric transition. To this
end, we have performed Density Functional (DFT) electronic structure 
calculations \cite{Hoh64,Koh65} for KDP, within a gradient corrected 
approach to exchange and correlation, both in the tetragonal paraelectric 
phase (forcing the protons to be in the center of the H-bonds) and in the 
orthorhombic ferroelectric phase, where the protons order cooperatively at 
off-center positions in the H-bond. We thus analyzed in detail the structural 
parameters and the microscopic electronic density redistributions associated 
to the proton ordering phenomenon, which eventually provoke the 
ferroelectric instability. 

This work is organized as follows: in Section 2 we describe the methods 
and the details of the calculations. Section 3 is devoted to the analysis 
of the geometry results. In Section 4, we study and discuss the 
polarization mechanism produced by the off-centering of the hydrogen atoms
along the H-bonds, and its relation to the ferroelectric instability. 
Finally, in Section 5 we elaborate our conclusions.

\section{Ab initio methods}

We have performed two different types of {\it ab initio} calculations. First,
we have used a recently developed, very efficient localized basis (LB) set 
approach (SIESTA). This is a fully self-consistent, pseudopotential 
DFT method that employs a basis set of pseudoatomic orbitals 
(PAO).\cite{Ord96,San97} 
We have chosen a double-zeta basis set with polarization functions (DZP). 
The exchange-correlation energy terms were computed using the 
Perdew-Burke-Ernzerhof gradient-corrected functional.\cite{Per96} 
Angular-dependent, norm-conserving Troullier-Martins 
pseudopotentials \cite{Tro91} were employed to represent the interaction 
between ionic cores and valence 
electrons. We have also included nonlinear core corrections (NLCC) for 
a proper description of the K ion, due to the large overlap of core 
and valence charge densities. We used a 125 Ry energy cutoff for the 
grid computation of numerical integrals \cite{San97,Ord96}, and
a value of $E_c=50$ meV for the orbital confinement energy, which gives total 
energies and geometries of sufficient accuracy for this system. 
Due to the approximate character of these confined basis orbitals, we have 
checked the overall accuracy by performing standard pseudopotential
plane wave (PW) calculations. The parallel version of the Car-Parrinello 
SISSA FPMD code \cite{Cav99}, and the same set of pseudopotentials, 
exchange-correlation functional, and NLCC were used. The energy cutoff for 
the plane wave expansion was set to 150 Ry. 
In both approaches, the electronic Brillouin zone sampling reduced to
the $\Gamma$-point, which is a reasonable approximation for such a
wide band gap insulator. We have, however, checked that this sampling is 
sufficient for the large supercells considered.
  
In the paraelectric phase, the KDP structure (space group I$\bar{4}$2d) 
has a body-centered tetragonal primitive lattice. For our calculations, we 
used instead the conventional face-centered tetragonal cell
(space group F$\bar{4}$d2),\cite{Bac55} which includes 64 atoms 
(8 formula units). With this choice, the transformation to the orthorhombic
structure of the ferroelectric phase is simply described by a change in the
ratio of the basal plane lattice parameters, in addition to changes in the
internal structure parameters.

\section{Geometry results} 

\begin{table}
\caption{\small Comparison of the ab-initio (LB and PW) lattice parameters and 
internal coordinates with experimental data. The notation is the same used 
in the experimental works referred. The different cases considered in the 
calculations are explained in section 2. We defined $\Delta z_K$ as the 
displacement of the K atom along z, respect to its centered position between P planes. 
Distances are in~\AA~and angles in degrees.}

\begin{center}
\begin{tabular}{  |l | c  c  c | c  c | c  c |}
\hline\hline
 &  &  LB  &  & PW &  & Exp. \cite{Tun88}  & Exp. \cite{Nel87} \\
 &  &      &  &  &     & Tetr.  & Orth. \\
 &TS& FP$_t$ & GS &TS& FP$_t$ & (293K) & (T$_c$-20K)\\
\hline\hline
Latt. param. & &         &          &        &        &         &         \\
 a   &  10.54  &  10.54  &   10.75  & 10.54  & 10.54  & 10.539  &  10.546 \\
 b   &  10.54  &  10.54  &   10.71  & 10.54  & 10.54  & 10.539  &  10.466 \\
 c   &   6.97  &   6.97  &    6.98  &  6.97  & 6.97   &  6.974  &   6.926 \\ 
\hline
Int. Coord.  & &         &          &        &        &         &         \\
d(P-O2) & 1.592 & 1.622  &  1.633   & 1.557   & 1.580  & 1.5403  & 1.5719  \\  
d(P-O1) & 1.592 & 1.572  &  1.573   & 1.557   & 1.515  & 1.5403  & 1.5158  \\  
2R      & 2.407 & 2.457  &  2.482   & 2.445   & 2.503  & 2.4946  & 2.4974  \\ 
$\delta$ & 0    & 0.259  &  0.311   & 0      & 0.395  & 0.3647  &  --  \\ 
$\Delta z_K$ & 0 & 0.07  &  0.09    & 0      & 0.09   & 0       & 0.102   \\ 
$<$ O2-P-O2 & 108.6 & 107.1 & 105.0 & 111.1  & 105.2  &  110.52 & 106.65  \\ 
$<$ O1-P-O1 & 108.6 & 114.7 & 114.1 & 111.1  & 115.1  &  110.52 & 114.80  \\ 
$<$ O1$\cdots$H-O2 & 176.1 & 176.5 & 177.1 & 180 & 179 & 177.23 & 179.44  \\
$\theta$    & 59.8  &  61.3 &  60.8 &  60.2  &  61.4  &   60.89 & 61.64   \\ 

\hline\hline



\end{tabular}
\end{center}
\vspace{-0.3truecm}
\end{table}

\par In order to obtain the equilibrium configurations relevant to 
ferroelectricity in KDP, we performed a series of computational 
experiments, which are described below.

First we considered a static view of the paraelectric phase by fixing a
tetragonal cell with parameters at their experimental values at 
293 K \cite{Nel87,Tun88} and relaxed the atomic positions under the
constraint that the H atoms remain in the middle of the H-bonds.
In this way we found the transition state (TS) which connects the two 
oppositely polarized realizations of the ferroelectric phase.
In a second time we still kept fixed the tetragonal lattice, and allowed the
hydrogen atoms to relax from their centered positions. When the unstable TS
state was slightly perturbed, two H atoms approach each PO$_4$ tetrahedron.
The two PO$_4$ oxygens which were approached by the hydrogens 
(in the following named O2) lie on one side of the (xy)-plane containing 
the P ion. The other two oxygens (O1) lie on the other side. The geometry
relaxation is carried out also for all the other atoms in the unit cell. 
As a consequence of proton ordering, also the K ions move away from the 
centered position between P planes, thus leading to a ferroelectric phase 
with tetragonal lattice, which we denote FP$_{t}$. 
Finally, we fully optimized atomic positions and cell parameters. This 
leads to an orthorhombic ground state struture GS, which corresponds 
with the actual experimental ferroelectric structure.

The resulting structural parameters are shown in 
Table 1, together with experimental data for the paraelectric tetragonal 
and the ferroelectric orthorhombic phases. An excellent overall agreement 
is observed for the TS and FP$_{t}$ configurations obtained with the PW 
method, as compared to experimental data. The good quality of the LB method
can be assessed by comparison with the PW results in columns 4 and 5.
In the FP$_{t}$ calculation, the off-center ordering of the hydrogen atoms
along the H-bonds leads to inequivalent O1 and O2 oxygens. Consequently, (a)
there is a distortion of the PO$_4$ tetrahedra (see the different P-O1 and
P-O2 distances on Table 1) and (b) the K ions abandon the
centered positions between the phosphate planes 
(see $\Delta z_K$ in Table 1), thus
inducing a spontaneous polarization which leads to ferroelectricity.
These effects will be explained in the next Section. We also verified that, 
when the hydrogen atoms are constrained to remain in the middle of the H-bonds, 
then the centered positions are stable for the K ions. This provides a 
strong evidence that the origin of ferroelectricity is in the off-centering
of the protons in the H-bonds.

It is worth mentioning that, in this fixed tetragonal FP$_{t}$  calculation,
a substantial anisotropic stress appears in the $xy$-plane. Note that the 
hydrogen off-centering in the H-bond network, which lies nearly on the 
$xy$-plane, leads to an asymmetry of the $x$ and $y$ directions. Further
relaxation, obtained by optimizing also the cell parameters in order to
minimize the stress with respect to ambient conditions, releases the 
tetragonal symmetry in favor of the orthorhombic structure shown in column
3 of Table 1.

\section{Polarization mechanism for the ferroelectric instability}

\begin{table}
\caption{\small Mulliken populations for each atom in the FP$_t$ and TS
cases for the LB calculation. The last line shows the charge
difference $\Delta q = q($FP$_t) - q($TS$)$ between both cases.
Units are electrons (e) for the first two rows, and e/1000 for 
the last one.}
\begin{center}
\begin{tabular}{  |l | c  c  c  c  c  |}
\hline\hline
                &     O1     &    O2    &    P    &   K    &    H    \\
\hline
FP$_t$          &   6.283    &  6.143   &  4.747  &  0.205 &  1.099  \\
TS              &   6.201    &  6.201   &  4.755  &  0.208 &  1.116  \\
$\Delta q/1000$ &   +82      &  -58     &    -8   &   -3   &  -17    \\
 
\hline\hline

\end{tabular}
\end{center}
\vspace{-0.5truecm}
\end{table}

In order to understand the origin of the ferroelectric instability,
we have focused our analysis in the electronic charge redistributions 
that happen in going from the centered (TS) to the off-centered
hydrogen case (FP$_t$). To this purpose we compute the orbital Mulliken 
populations, which are shown on Table 2 for the different atoms in both 
cases. It is known that Mulliken populations
depend strongly on the choice of the basis set. Differences, however, 
are much less sensitive, and thus meaningful. In going from TS to FP$_t$, 
it can be clearly observed an increase of the charge 
localized around O1, the main contribution ($\approx 70$\%) being
provided by the O2 charge decrease.
The analysis of bond and self atom-orbitals
overlap populations,\cite{Mul55} which are shown in
Table 3, also helps to discriminate the origin of this charge redistribution.

The trends observed in Tables 2 and 3 are confirmed by the charge density 
difference $\Delta \rho({\bf r})=\rho_{FP_t} ({\bf r})-\rho_{TS}({\bf r})$, 
which is plotted in the planes determined by the atoms P-O2-H (Fig. 1a) and 
P-O1$\cdots$H (Fig. 1b). Analysing both, Table 3 and Fig. 1a, we observe
a significant enhancement of the O1-O1 overlap population, accompanied by a 
smaller increment of the same sign in the O1-P orbitals. This happens at the
expenses of the population of the overlap orbitals in the O1$\cdots$H bond.
This behaviour around the O1 atom indicates that, as the hydrogen atoms move 
away from O1, the strongly covalent O1-H bond weakens, its charge being
redistributed. A substantial amount of this charge localizes around O1, 
while part of it localizes in the O1-P orbitals, consistently with the 
shortening of the O1-P bonds reported in Table 1.
The contrary occurs in the vicinity of the O2 atom: as H moves towards O2 
along the O1$\cdots$H-O2 line, the O2-H distance decreases, thus
increasing the degree of covalency and producing 
a charge localization along the corresponding O2-H bond. This is 
indicated by the bond overlap population in Table 3, 
and also by the contours in Fig. 1b. Finally, the O2-O2 and O2-P orbitals
loose a significant amount of charge.
We name this process of charge redistribution, produced by the collective 
hydrogen off-centering along the H-bond, the polarization mechanism (PM). 
The PM can be described in a simplified way as a charge accumulation and 
localization around O1 at the expenses of a charge depletion and 
delocalization in the vicinity of O2. 

This charge accumulation around the O1 atom can also be viewed as a negative 
charge defect, which electrostatically attracts the positive 
K$^+$ ion towards the PO$_4$ cage, along the $z$-axis. This electrostatic 
force is counterbalanced by repulsion forces from orbital overlaps, which 
stabilize a new equilibrium position of the K$^+$ ion, as can also be seen 
in Table 1. Our calculation then supports the idea that proton off-centering 
and ferroelectricity are very correlated phenomena.

\begin{table}
\caption{\small Overlap Populations between atoms in units of e/1000.
Only the most significant $\Delta q = q($FP$_t) - q($TS$)$ values
are shown.}
\begin{center}
\begin{tabular}{  |l | c  c  c |  c  c  c | c |}
\hline\hline
  &   O1-O1    &   O1$\cdots$H   &  O1-P   &  O2-O2  &  O2-H &  O2-P  &   P-P  \\
\hline
$\Delta q/1000$  &   +200    &  -91   &    46   &  -92  &  70  & -44 &  -23  \\
\hline\hline
\end{tabular}
\end{center}
\vspace{-0.5truecm}
\end{table}

\begin{figure}[thb]
\psfig{figure={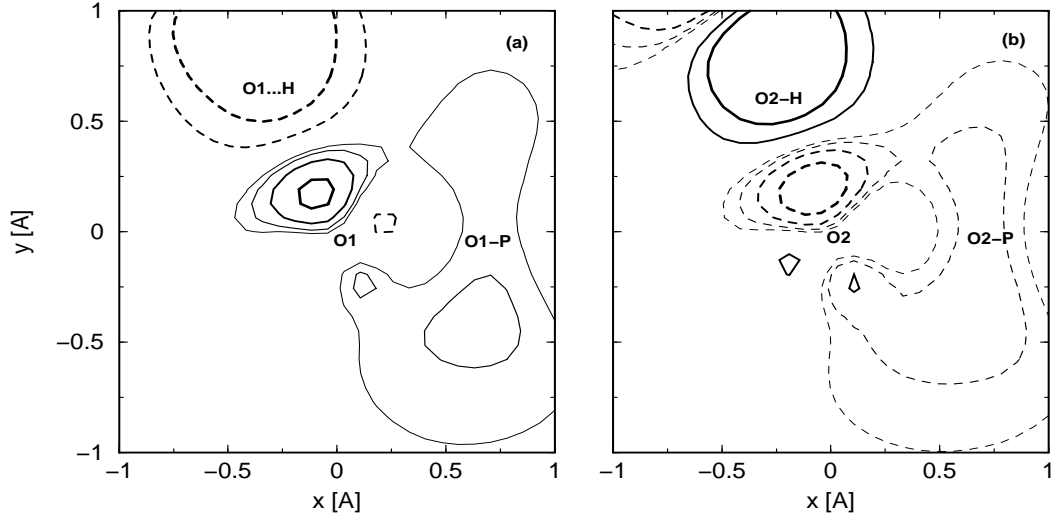},angle=270,width=14cm,height=7cm}
\vspace{-0.25truecm}
\caption{\small Differential charge density contours $\Delta \rho({\bf r})$
in the planes containing the following atoms: (a) P-O1$\cdots$H,  
(b) P-O2-H. Labels O1 and O2 denote the positions of the respective
nuclei, positioned at (0,0). Labels O2-P and O1-P indicate the position 
of the center of the corresponding bonds. The same convention is used
for the O2-H  and O1$\cdots$H bonds. Positive (negative) contours are in
solid (dashed) lines.
The thickest lines represent an absolute value of $2.96 \times 10^{-3}$ 
e\AA$^{-3}$. The thinner lines are obtained by successively halving this 
value, down to $3.70 \times 10^{-4}$ e\AA$^{-3}$.}
\end{figure}

Cell relaxation -- from FP$_t$ to GS -- also affects the internal 
structure (Table 1). As H approaches O2, $\delta$ increases, the O1$\cdots$H 
bond weakens and the H-O2 bond becomes stronger. 
As a consequence, a larger amount of charge localizes around O1, and
delocalizes away from O2. This behavior is related to the PM and will be
useful to establish a connection between $\delta$ and the charge 
difference $\Delta q_O = q(O1) - q(O2)$. Moreover, we observe a 
striking correlation between $\Delta q_O$, the distortion of the 
tetrahedra, $\Delta$(P-O) $\equiv d(P-O2)-d(P-O1)$, and the displacement 
of the K atoms, $\Delta z_K$. These trends are shown in the theoretical curves 
in Fig. 2 for two different values of $\delta$, obtained from the 
cell-unrelaxed FP$_t$ and the GS calculations. 

The crucial experimental observation that the effect of isotopic substitution 
can be almost exactly reproduced by applying pressure \cite{Nel87,McM90}, 
supports the idea that the main effect of deuteration is to modify the internal 
geometry. In this sense, our calculations for FP$_t$ and GS, corresponding to 
two different pressures, can be used to analyse isotopic effects.
It is known from experiment that H(D) ordering, 
$\Delta z_K$, $\Delta$(P-O), and spontaneous polarization P$^s$, show 
very similar trends just below T$_c$, in the ferroelectric phase.\cite{Nel85}
Moreover, these trends are quite well explained by the present PM (left curves in Fig. 2). 
Therefore, it is natural to attempt to establish a connection between 
$\Delta q_O$ and structural data measured in KDP and in its deuterated analog,
DKDP. This is done in Fig. 2, by including the experimental values of 
$\Delta z_K$, $\Delta$(P-O) and P$^s$ for KDP and deuterated DKDP. The 
comparison between theoretical and experimental curves is very favorable,
especially for what concerns the slopes of the curves.\cite{disclaimer} 

In connection with the proposed PM, the above indicates that the enhancement 
of the saturated polarization due to deuteration, 
experimentally observed below T$_c$, \cite{Nel85} can be explained 
merely by the increase in the charge unbalance between O1 and O2 
($\Delta q_O$), which in turn is induced by an increase in $\delta$. 
This picture is in line with the ideas developed in Ref.\cite{Koj88}, 
where the experimental behaviour of P$^s$ and the inverse dielectric 
constant were rationalized in terms of a proton-dipole interaction model. 
In fact, at temperatures well below T$_c$, the protons are strongly bound 
to the neighboring PO$_4$ dipoles, in a stable minimum of an asymmetric 
potential. In this asymmetric configuration, the protons localize into the 
deepest well because the energy needed to be promoted to the excited state is
too high, thus ruling out the possibility of tunnelling.\cite{Koj88} 
In this case, deuteration can only induce the indirect effect of enhancing 
$\delta$ \cite{Mat82}, {\it i.e.} the so-called geometric effect. Preliminary 
{\it ab initio} calculations indicate, even more strikingly, that there is 
actually no double minimum structure in the potential energy surface for single 
protons. Appart from supporting this scenario, the current 
{\it ab initio}-derived PM establishes the importance of the short-range
proton~--~PO$_4$-dipole interaction for the transition, as was also speculated in 
\cite{McM90}. 
  
\vspace{0.2truecm}
\begin{figure}[thb]
\centerline{\epsfxsize=3.0truein
\psfig{figure={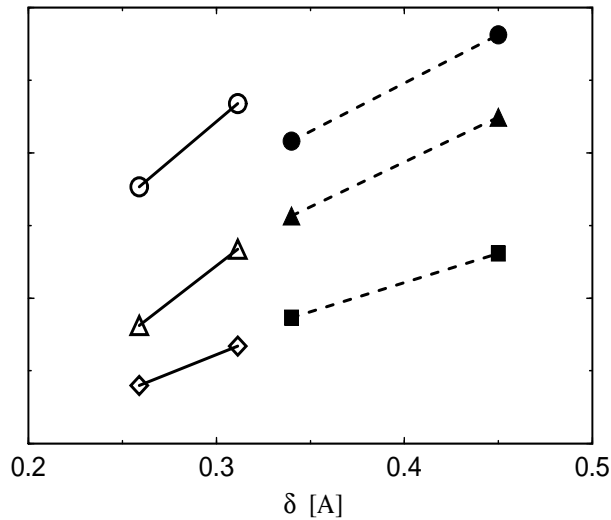},angle=270,width=8cm,height=7cm}}
\vspace{-0.3truecm}
\caption{\small The variaton of $\Delta$(P-O) (circles), $\Delta z_K$ (triangles),
$\Delta q_O$ (diamonds) and saturated P$^s$ (squares) with the
H(D) displacement $\delta$. Left curves: LB theoretical results for
cases FP$_t$ and GS (open symbols) and right curves: experimental
values for KDP with $\delta^{exp}_H \approx 0.34 \AA$, and for DKDP
with $\delta^{exp}_D \approx 0.45 \AA$, from Ref.\cite{Nel85}, (full symbols),
both in the ferroelectric phases at T$_c$-20K and T$_c$-10K, respectively. 
Experimental $\delta$ values considered are estimations of these magnitudes at T$_c$,
as they cannot be measured in the ferroelectric phase.\cite{Nel85}
The lines are guides to the eye only. Arbitrary units are used in the vertical axis.}
\vspace{-0.5truecm}
\end{figure}

\section{Conclusions}

We presented the results of {\it ab initio} DFT 
calculations for the H-bonded ferroelectric material KDP, which reproduce 
very well the structure of the tetragonal and orthorhombic 
(ferroelectric) phases in KDP. On this basis we identified the 
microscopic polarization mechanism that leads to the ferroelectric 
instability. This arises from an electronic charge redistribution between
the oxygen atoms involved in the H-bonds, occurring when the protons move 
away from the center of the bonds. We indirectly showed that the interplay 
between proton ordering and ferroelectric polarization qualitatively explains 
the enhancement of the saturated P$^s$ upon deuteration. In addition we provided, 
for the first time, fully {\it ab initio} evidence supporting the idea that 
isotope effects reduce, at least at low temperatures in the ferroelectric phase, 
to the {\it indirect} geometric effect of enhancing $\delta$, as speculated in 
recent works \cite{McM90}.

\section*{Acknowledgments}

We thank E. Tosatti for his invaluable collaboration in the 
initial stages of this work. JK also acknowledges very useful discussions 
with R.J. Nelmes and M.I. McMahon.
We also thank P. Ordej\'on for his assistance in the use 
of the SIESTA package and C. Cavazzoni for providing us the FPMD SISSA code.
RLM and SK acknowledge support from ICTP (Italy) and CONICET (Argentina).


\begin{thebibliography}{99}
\bibitem{disclaimer} Notice that our calculations actually correspond to 
                     infinitely massive nuclei, and also dynamical aspects are 
                     not taken into account.
\end{thebibliography}
\end{document}